\documentclass[useAMS,usenatbib]{mn2e}
\usepackage[dvips]{graphicx}
\usepackage[english]{babel}
\usepackage[latin1]{inputenc}
\usepackage{latexsym}
\usepackage{rotating}
\usepackage{amssymb,amsmath}

\def\bbbc{{\mathchoice {\setbox0=\hbox{$\displaystyle\rm C$}\hbox{\hbox
to0pt{\kern0.4\wd0\vrule height0.9\ht0\hss}\box0}}
{\setbox0=\hbox{$\textstyle\rm C$}\hbox{\hbox
to0pt{\kern0.4\wd0\vrule height0.9\ht0\hss}\box0}}
{\setbox0=\hbox{$\scriptstyle\rm C$}\hbox{\hbox
to0pt{\kern0.4\wd0\vrule height0.9\ht0\hss}\box0}}
{\setbox0=\hbox{$\scriptscriptstyle\rm C$}\hbox{\hbox
to0pt{\kern0.4\wd0\vrule height0.9\ht0\hss}\box0}}}}
\def\bbbq{{\mathchoice {\setbox0=\hbox{$\displaystyle\rm
Q$}\hbox{\raise
0.15\ht0\hbox to0pt{\kern0.4\wd0\vrule height0.8\ht0\hss}\box0}}
{\setbox0=\hbox{$\textstyle\rm Q$}\hbox{\raise
0.15\ht0\hbox to0pt{\kern0.4\wd0\vrule height0.8\ht0\hss}\box0}}
{\setbox0=\hbox{$\scriptstyle\rm Q$}\hbox{\raise
0.15\ht0\hbox to0pt{\kern0.4\wd0\vrule height0.7\ht0\hss}\box0}}
{\setbox0=\hbox{$\scriptscriptstyle\rm Q$}\hbox{\raise
0.15\ht0\hbox to0pt{\kern0.4\wd0\vrule height0.7\ht0\hss}\box0}}}}
\def\bbbt{{\mathchoice {\setbox0=\hbox{$\displaystyle\rm
T$}\hbox{\hbox to0pt{\kern0.3\wd0\vrule height0.9\ht0\hss}\box0}}
{\setbox0=\hbox{$\textstyle\rm T$}\hbox{\hbox
to0pt{\kern0.3\wd0\vrule height0.9\ht0\hss}\box0}}
{\setbox0=\hbox{$\scriptstyle\rm T$}\hbox{\hbox
to0pt{\kern0.3\wd0\vrule height0.9\ht0\hss}\box0}}
{\setbox0=\hbox{$\scriptscriptstyle\rm T$}\hbox{\hbox
to0pt{\kern0.3\wd0\vrule height0.9\ht0\hss}\box0}}}}
\def\bbbz{{\mathchoice {\hbox{$\sf\textstyle Z\kern-0.4em Z$}}
{\hbox{$\sf\textstyle Z\kern-0.4em Z$}}
{\hbox{$\sf\scriptstyle Z\kern-0.3em Z$}}
{\hbox{$\sf\scriptscriptstyle Z\kern-0.2em Z$}}}}

\newcommand{\beq}{\begin{equation}}
\newcommand{\beqa}{\begin{eqnarray*}}
\newcommand{\beqan}{\begin{eqnarray}}
\newcommand{\greq}{\begin{equation}\left\{ \begin{array}{l}}
\newcommand{\egreq}{\end{array}\right. \end{equation}}
\newcommand{\nngreq}{\[\left\{ \begin{array}{l}}
\newcommand{\nnegreq}{\end{array}\right. \]}
\newcommand{\egreqn}[1]{\end{array}\right. \label{#1}\end{equation}}
\newcommand{\eeq}{\end{equation}} 
\newcommand{\eeqn}[1]{\label{#1}\end{equation}} 
\newcommand{\eeqa}{\end{eqnarray*}}
\newcommand{\eeqan}[1]{\label{#1}\end{eqnarray}}

\newcommand{\tv}{ \rightarrow}

\newcommand{\lp}{ \left(}
\newcommand{\rp}{ \right)}
\newcommand{\lc}{ \left[}
\newcommand{\rc}{ \right]}
\renewcommand{\la}{ \left\{}
\newcommand{\ra}{ \right\}}

\newcommand{\llp}{\ell(\ell+1)}

\newcommand{\iml}{ \frac{im}{\ell(\ell+1)} }

\newcommand{\na}{ \vec{\nabla} }

\newcommand{\vu}{\vec{u}}

\newcommand{\er}{\vec{e}_r}

\newcommand{\etheta}{\vec{e}_\theta}

\newcommand{\ephi}{\vec{e}_\varphi}

\newcommand{\ez}{\vec{e}_z}

\newcommand{\vb}{\vec{b}}
\newcommand{\vB}{\vec{B}}

\newcommand{\vzero}{\vec{0}}

\newcommand{\od}[1]{\mbox{${\cal O}(#1)$}}

\newcommand{\dnr}[1]{\frac{d  #1}{dr}}

\newcommand{\ddnr}[1]{\frac{d^2  #1}{dr^2}}

\newcommand{\dz}[1]{\frac{\partial  #1}{\partial z}}

\newcommand{\dxx}{\frac{\partial}{\partial x}}

\newcommand{\eq}[1]{(\ref{#1})}

\newcounter{sub}
\newcounter{subeqn}[sub]
\renewcommand{\thesubeqn}{\alph{subeqn}}
\renewcommand{\theequation}{\thesub\thesubeqn}

\newcommand\OOmega{\mbox{\boldmath $\Omega$}}

\newcommand\nab{\mbox{\boldmath $\nabla$}}

\bibpunct{(}{)}{;}{a}{}{,}%
\newcommand{\e}{\mathrm{e}}

\renewcommand {\ephi}{{\bf e}_{\varphi}}
\renewcommand {\er}{{\bf e}_r}
\newcommand {\x}{{\bf x}}

\renewcommand {\ez}{{\bf e}_z}

\newcommand {\rot}{{\nab}\*}
\newcommand {\vect}{\bf }
\newcommand{\allp}[2]{A^{#1}_{#2}}
\newcommand{\allm}{\alpha^{\ell}_{\ell-1}}
\renewcommand{\allp}{\alpha^{\ell}_{\ell+1}}
\newcommand{\Allm}{A^{\ell}_{\ell-1}}
\newcommand{\Allp}{A^{\ell}_{\ell+1}}
\newcommand{\Bllm}{B^{\ell}_{\ell-1}}
\newcommand{\Bllp}{B^{\ell}_{\ell+1}}
\newcommand{\almm}{a^{\ell-1}_m}
\newcommand{\almp}{a^{\ell+1}_m}
\newcommand{\alm}{a^{\ell}_m}

\newcommand{\blm}{b^{\ell}_m}

\newcommand{\clmm}{c^{\ell-1}_m}
\newcommand{\clmp}{c^{\ell+1}_m}
\newcommand{\clm}{c^{\ell}_m}

\newcommand{\dpart}[2]{\frac{\partial #1}{\partial #2}}
\newcommand{\dddnr}[1]{\frac{d^3  #1}{dr^3}}

\newcommand{\Tll}{\vect{T}_{\l}^\l}

\newcommand{\ulmm}{u^{\ell-1}_m}
\newcommand{\ulmp}{u^{\ell+1}_m}
\newcommand{\ulm}{u^{\ell}_m}
\newcommand{\vlmm}{v^{\ell-1}_m}
\newcommand{\vlmp}{v^{\ell+1}_m}
\newcommand{\vlm}{v^{\ell}_m}
\newcommand{\wlmm}{w^{\ell-1}_m}
\newcommand{\wlmp}{w^{\ell+1}_m}
\newcommand{\wlm}{w^{\ell}_m}
\newcommand{\wkm}{w^{k}_m}

\renewcommand {\*} {\times}
\renewcommand {\.} {\cdot}

\renewcommand {\div}{{\bf \nab}\.}
\renewcommand {\l}{\ell}

\def\lp{\left(}
\def\rp{\right)}

\def\st{\stepcounter{sub}}
\def\stq{\stepcounter{subeqn}}
\def\be{\begin{equation}}
\def\ee{\end{equation}}
\def\bea{\begin{eqnarray}}
\def\eea{\end{eqnarray}}
\def\bean{\begin{eqnarray*}}
\def\eean{\end{eqnarray*}}

\newcommand{\LE}{\mathrm{Le}}

\def\b{{\bf b}}

\def\u{{\bf u}}
\def\v{{\bf v}}
\def\r{{\bf r}}

\newcommand\B{{\bf B}}

\newcommand\no{\nonumber}
\begin{document}

\title[]{A $r$-mode in a magnetic rotating spherical layer: \\ application to
neutron stars.}
\author[S. Abbassi, M. Rieutord, and V. Rezania]{S. Abbassi$^{1,2}$\thanks{E-mail:
abbassi@ipm.ir}
, M. Rieutord$^{3,4}$\thanks{E-mail:rieutord@ast.obs-mip.fr} and V. Rezania$^5$
\thanks{Email:rezani
av@macewan.ca}\\
$^{1}$School of Physics, Damghan University of Basic Sciences,
 P.O.Box 36715-364, Damghan, Iran\\
$^{2}$School of Astronomy, Institute for Studies in Theoritical Physics and Mathematics,
P.O.Box 193
95-5531, Tehran, Iran\\
$^{3}$Universit\'e de Toulouse; UPS-OMP; IRAP; Toulouse, France\\
$^{4}$CNRS; IRAP; 14, avenue Edouard Belin, F-31400 Toulouse, France\\
$^{5}$Department Physical Sciences, Grant Macewan University, Edmonton T5J 4S2, Canada}

\date{\today}

\pagerange{\pageref{firstpage}--\pageref{lastpage}} \pubyear{2011}

\maketitle \label{firstpage}
\begin{abstract}
The impact of the combination of rotation and magnetic fields on
oscillations of stellar fluids is still not well known theoretically. It
mixes Alfv\'en and inertial waves. Neutron stars are a place where both
effects may be at work. We wish to decipher the solution of this problem
in the context of $r$-modes instability in neutron stars, as it appears
when these modes are coupled to gravitational radiation.

We consider a rotating spherical shell filled with a viscous fluid but
of infinite electrical conductivity and analyze propagation of modal
perturbations when a dipolar magnetic field is bathing the fluid layer. We
perform an extensive numerical analysis and find that the $m=2$ $r$-mode
oscillation is influenced by the magnetic field when the Lehnert number
(ratio of Alfv\'en speed to rotation speed) exceeds a value proportional
to the one-fourth power of the Ekman number (non-dimensional measure
of viscosity).  This scaling is interpreted as the coincidence of the
width of internal shear layers of inertial modes and the wavelength of the
Alfv\'en waves.  Applied to the case of rotating magnetic neutron stars,
we find that dipolar magnetic fields above $10^{14}$ G are necessary to
perturb the $r$-modes instability.

\end{abstract}

\section{Introduction}

Numerous astrophysical systems exhibit a pulsating behavior that can
be significantly influenced by both magnetic field and rotation. The
rapidly oscillating Ap (roAp) stars, the magnetic white dwarf stars and
neutron stars as well as planetary cores fall into the above category.

In neutron stars, however, possible astrophysical implications of
$r$-modes instability have motivated extensive investigations of this mode
during the past few years (see \citealt{And03} for a review). $r$-modes
belong to the class of inertial modes that arise in rotating fluids due
to the Coriolis force. It was shown by \cite{anderss98} and \cite{FM98}
that these modes easily couple to gravitational radiation and become
unstable, allowing the neutron stars to lose their angular momentum.

The foregoing instability may however be weakened or even suppressed by
all the dissipative mechanisms which couple to an $r$-mode oscillations.
Much work has thus been devoted to the analysis of the various mechanisms
which may damp $r$-modes. Recently, vortex-mediated mutual friction of
superfluids was investigated by \cite{HAP09} as well as the effects of
hyperon bulk viscosity \cite[][]{HA10}, but both effects do not seem to
be able to influence the instability at low temperatures. For these
temperatures, the Ekman layer which forms below
the crust of a neutron star, remains the most important
source of dissipation \citep{BU00,R01,GA06}.

However, it is well-known that the magnetic field is an important
component of a neutron star. It is therefore clear that fluid flows may be
seriously influenced by this field and that other channels of dissipation
for the $r$-mode instability may exist through this component. Much work
was devoted to investigate the modifications induced by a magnetic field
on these kind of modes. \cite{Rez00,Rez01a,Rez01b}, for instance, showed
that strong magnetic field, beyond $10^{10}$G may weaken the $r$-mode
instability sufficiently so as to make the generated gravitational waves
undetectable. \cite{Men01} and \cite{KM03} focused their study
on the influence of the magnetic field on the Ekman layer flow. They
conclude that magnetic field larger than $10^{12}$G completely suppress
the instability. However, \cite{Lee05}, using a dipolar magnetic field
covering the surface of a neutron star modeled as a $N=1$-polytrope,
concludes that much stronger magnetic fields, over $10^{14}$G, are
necessary  to suppress the instability through magnetic perturbations.
Let us note that beside the suppression of the foregoing instability,
magnetic fields can also directly spin down a neutron star through the
classical process of magnetic braking \cite[i.e.][]{HL00}. More recently,
\cite{CD10} and \cite{BCDPS11} investigated the role of magnetic field
generation by the unstable $r$-mode, showing that a very strong toroidal
magnetic field can be generated, which in turn can modify the instability.

The aim of the present paper is to explore furthermore the channels
of dissipation for the unstable $m=2$ $r$-mode of a rotating neutron
star, especially when viscosity and magnetic fields are both present.
Indeed, a possibility that has not been considered by previous studies
(like the one of \citealt{Lee05}), is the fact that an unstable $r$-mode
of a spherical layer, might develop internal shear layers thanks to
the magnetic field action. Since fluid layers of different natures are
expected due to phase transitions of nuclear matter, the existence of
internal shear layers are also very likely. We shall see that this leads
to a threshold value around $10^{14}$G for magnetic fields to noticeably
perturb the $r$-mode instability.

Rotating fluid layers bathed by a magnetic field are however not specific
to neutron stars. They are also found in planetary core, like in the
Earth's core.  Thus, in order to make the following study of more general
interest, we shall consider a very simplified model of neutron star,
neglecting relativistic or superconducting effects. We thus extend the
results of \cite{R01} to the case where a dipolar magnetic field perturb
the fluid flow (in the limit of very large magnetic Prandtl numbers). This
is meant to be a simple configuration where combined effects of magnetic
fields and viscosity can be studied.

The paper is organized as follows: In Sect. 2, we recall the basic
physical ingredients of the model and then (sect. 3) briefly explain
the numerical strategy. Numerical results for the $m=2$ $r$-mode coupled
with Alfv\'en waves are discussed in Sect. 4. Conclusions follow.

\section{The model}

We consider a rotating star modeled as an infinitely electrically
conducting core surrounded by a spherical layer of fluid itself limited
by an outer solid crust. The ratio of the inner core radius to the outer
radius $R$ is $\eta$. The kinematic viscosity and magnetic diffusivity
of the fluid are respectively $\nu$ and $\nu_m$.

The star is rotating with uniform angular frequency $\OOmega=\Omega \ez$
along the z-axis.  The core is supposed to be the source of a permanent
axisymmetric magnetic dipole covering the whole layer.  The symmetry
axis of the magnetic field is along the z-axis.  Note that such a
partition of the star is necessary in order to avoid the problem of the
definition of the magnetic field in the core of the star, the dipole
field being singular at the center.   The fluid in the layer is taken to
be incompressible; we thus eliminate phenomena related to compressibility
such as $p$-modes (rapid and slow magnetic waves, when a magnetic field
is applied). Classical MHD approximations are used (no charge separation,
non-relativistic motions, etc.).

\subsection{Equations of motion}
The shell is bathed by an axisymmetric dipolar magnetic field:
\st
\begin{equation}
\label{dipole}
   \B_0 = B_p \. R^3 \left( \frac{\cos \theta}{r^3} \er
         + \frac{\sin \theta} {2r^3} \etheta \right),
\end{equation}
where ($\er,\etheta,\ephi$) are spherical coordinate unit vectors. $B_p$ is the
amplitude of the magnetic field at the surface poles of the star.

The equations of motion for an incompressible rotating fluid can be written as
\st
\bea\label{eqm}
\stq\label{euler}
&& \dpart{\v}{t} +  \v\cdot\nab\v+ 2 \OOmega\*\v = \nonumber\\
&&  - \nab p/\rho  - \nab\Phi_{\rm eff} +
  \frac{1}{\mu_0\rho}(\rot \B) \* \B + \nu\Delta \v , \\
\stq\label{rotB}
&&\dpart{\B}{t} = \rot \left(\v \* \B \right) + \nu_m \Delta\B\\
\stq\label{divB}
&& \div \B = 0,\\
\stq\label{divV}
 &&\div \v =0,
\eea
where $\Phi_{\rm eff}=\Phi-\Omega^2 r^2 \sin^2\theta/2$ is the
effective gravito-rotational potential and $\mu_0$ is the permeability of
vacuum.
Here $\v$ is the velocity field, $\B$ is the total magnetic
field (dipole field plus perturbation), and $\nu_m=1/(\mu_0\sigma)$ where
$\sigma$ is the fluid's electrical conductivity.

A non-dimensional form of the previous equations can be
obtained by introducing the parameters
\st
\[
V_{\rm A}=B_p/\sqrt{\mu_0\rho}, \quad V_\Omega= 2\Omega R, \quad
\LE=V_{\rm A}/V_\Omega,\]
\[ E=\nu/(RV_\Omega) , \quad E_m=\nu_m/(RV_\Omega),
\]
where $V_{\rm A}$ is the Alfv\'en speed, $V_\Omega$ is rotational speed, $E$
is the Ekman number and $E_m$ the magnetic Ekman number.
Following
\cite{Jault08}, we introduce $Le$, the Lehnert number in reference to the work of
\cite{Lehnert54}. As shown by its definition, this number measures the
ratio of the Alfv\'en speed to the rotation speed.

By linearizing magneto-hydrodynamic equations (\ref{eqm}) to the
kinetic $\u$ and magnetic $\b$ perturbations, one can study infinitesimal
perturbations from the equilibrium.  We assume these perturbations have a
time dependence of the form $\e^{\lambda t}$, where $\lambda =
\tau + i \omega$ ($\tau$ is the damping rate, $\omega$ the
pulsation and $i^2=-1$), and we use the following non-dimensional
variables:
\st
\be
\begin{array}{lll}
\v &\rightarrow& V_\Omega \u(\r)e^{\lambda t},\\
\B &\rightarrow& B_p \B(\r)+ B_p\b(\r)e^{\lambda t},\\
\rho &\rightarrow& \rho_o,
\end{array}
\ee
where $\u(\r)$ and $\b(\r)$ are now first order non-dimensional
quantities.
Therefore, Eqs. (\ref{eqm}) reduce to the following set of
equations:
\st
\bea
\label{eqm_pert}
\stq
       && \lambda \rot \u + \rot ( \ez \times \u)  =\nonumber \\
        && \qquad \LE^2 \rot \left( (\rot \b) \times \B
                \right) + E \rot \Delta \u, \\
\stq
      &&  \lambda \b = \rot \left(\u \times
                \B \right) + E_m \Delta \b,\\
\stq
       && \div \u = 0, \\
\stq
       && \div \b =0,
\eea
where $\B$ denotes the permanent dipolar magnetic field and we used $\rot\B=0$.
Here we take the curl of momentum equation in order to eliminate pressure
term.

\subsection{Boundary conditions}

Six boundary conditions are required in order to solve
Eqs. (\ref{eqm_pert}) uniquely. On the inner boundary $r=\eta R
~(\eta<1)$, the magnetic field perturbations have only tangent components
 because the core is assumed to be infinitely conducting. On the surface
$r=R$, the total magnetic field vector matches the external field which
is dipolar potential, as there are no currents in the external vacuum.
As for the velocity field, we may either use stress-free or no-slip
boundary conditions.

As for the magnetic field, different conditions apply for the inner and
outer boundaries. On the interior, the perturbation to the electric field
is perpendicular to the conducting core, and the perturbation to the
magnetic field is tangent.  This gives the following three equations:

\st
\bea
\stq
\label{b_bound}
         b_r &=& 0, \\
\stq
\label{b_bounda}
        \displaystyle
        \frac{E_m}{r} \dpart{}{r}(rb_{\theta})
                &=& -u_{\theta} B_r, \\
\stq
\label{b_boundb}
        \displaystyle
        \frac{E_m}{r} \dpart{}{r}(rb_{\varphi})
                &=& -u_{\varphi}B_r.
\eea
But actually only \eq{b_bounda} or \eq{b_boundb} together with
\eq{b_bound} is needed \cite[see][]{RRR04}.

The magnetic field perturbations outside the star ($r>R$) are derived from
a potential that does not diverge at infinity:
\st
\be\label{b_bound1}
\b_{\rm ext}=\nab \phi,
\ee
The boundary conditions at the surface of the star are just the continuity of the
magnetic field there. These conditions are easily
expressed after expansion of the fields in spherical harmonics (see appendix).

Eq.~(\ref{eqm_pert}), together with boundary conditions, defines a
generalized eigenvalue problem, where $\lambda$ is the eigenvalue and
$(\u, \b)$ is the eigenvector which can be computed numerically.

\subsection{The case of neutron stars}\label{ns_num}

In neutron stars, typical values for the various physical quantities are

\st
\bea
\stq
&&B_p\sim 10^{12}~{\rm G},\quad \rho\sim 10^{17}~ {\rm kg~m}^{-3},
\quad R\sim 10~{\rm km},\\
\stq
&& \Omega=2\pi\nu_s \sim 1900 ~{\rm rad~ s}^{-1}\\
\stq
&&\sigma\sim 5.2\times 10^{17} ~ {\rm ohm}^{-1}{\rm m}^{-1},\quad
\nu \sim 0.32 ~{\rm m}^{2}{\rm s}^{-1}, \\
\stq
&& \nu_m\sim 1.5 \times 10^{-12}~{\rm m}^{2}{\rm s}^{-1}.
\eea
Here, the values for $\sigma$ and $\nu$ are given  for temperature $T\sim
10^8$ K \citep{Bay69,FI76,FI79}.  Therefore, $V_{\rm A} \sim 2.8\times
10^2 $ m s$^{-1}$, $V_\Omega \sim 3.8\times 10^7$ m s$^{-1}$, so that

\[ \LE\sim
7.5\times 10^{-6}, \quad E\sim 8.4 \times 10^{-13}, \quad E_m\sim 4\times
10^{-24}.\]
We note that the magnetic Prandtl number $\nu/\nu_m=E/E_m\sim
10^{11}$ is extremely large. This means that the diffusion of magnetic
perturbations is a negligible source of dissipation. As a consequence,
we may simplify the set of equations \eq{eqm_pert} by setting $E_m=0$.
In this case the magnetic perturbation is readily given by the fluid
flow, namely

\[ \lambda\vb = \rot(\vu\times\vB)\]
Hence, the set of equations reduces to

\st
\bea
\label{eqm_perts}
\stq
       && \lambda \rot \u + \rot ( \ez \times \u)  =\nonumber \\
        && \frac{\LE^2}{\lambda} \rot \left[ (\rot \rot(\vu\times\vB)) \times \B
                \right] + E \rot \Delta \u, \\
\stq
       && \div \u = 0,
\eea
This system is completed by boundary conditions on the velocity
solely. Indeed, the magnetic field is completely frozen in the fluid and
the magnetic perturbations just represent the oscillations of the dipole
field lines. Interestingly, boundary conditions on the magnetic field
\eq{b_bound} show that if $E_m=0$ then $u_\theta=u_\varphi=0$. This means
that on the inner core boundary, no slip boundary conditions can be applied. This
is expected since the core is assumed at rest and no motion of the field
lines is allowed. We also assume no-slip boundary conditions ($\u=\vzero$)
on the outer boundary to take the crust into account.

The small value of the Lehnert number suggest that the coupling between
the $r$-modes and the magnetic field is quite weak. We readily see
from the simplified perturbations equations \eq{eqm_perts} that the
influence of Lorentz force will be noticeable compared to the viscous
one, if $\LE\geq\sqrt{E}$. Noting that the Lorentz operator and the
viscous operator are both of second order, we observe that no length
scale comes into this inequality. This means that it is valid both in
the Ekman boundary layers and in the bulk of the layer. From the numbers
that characterize neutron stars it is obvious that the inequality is met.  This 
means that the magnetic field influences the flow more than the viscosity
and possibly changes the instability of the $r$-modes. A more detailed
calculation is therefore necessary to assess the effect. Actually, we shall see
that the previous inequality is not stringent enough and a larger Lehnert
number is necessary to affect the instability.
We now turn to a numerical study of this problem.

\section{Numerical Method}

\subsection{Spherical harmonic projection}

To solve the eigenvalue problem expressed by Eq. (\ref{eqm_pert}), we
project the set of equations on the spherical harmonics in a similar
way as in \cite{R87,R91}. We expand the perturbed velocity and
magnetic fields into poloidal and toroidal components:

\st
\bea\label{eq:harmonic_decomposition}
\stq
    \v = \sum_{\l=0}^{\infty}
    \sum_{m=-\l}^{\l} u_m^{\l}(r) {\bf R}_{\l}^m  + v_m^{\l}(r) {\bf S}_{\l}^m
    + w_m^{\l}(r) {\bf T}_{\l}^m,\\
\stq
    \b = \sum_{\l=0}^{\infty}
    \sum_{m=-\l}^{\l} a_m^{\l}(r) {\bf R}_{\l}^m
    + b_m^{\l}(r) {\bf S}_{\l}^m
    + c_m^{\l}(r) {\bf T}_{\l}^m,
\eea
where the radial functions $\{u_m^\l,v_m^\l\}$ ($\{a_m^\l,b_m^\l\}$)
and $\{w_m^\l\}$ ($\{c_m^\l\}$) are the poloidal and toroidal parts of
the velocity (magnetic) fields, respectively.  ${\bf R}_{\l}^m$, ${\bf
S}_{\l}^m$, and ${\bf T}_{\l}^m$ are the vectorial spherical harmonics

\st
\be
{\bf R}_{\l}^m = Y_{\l}^m \er, \quad
{\bf S}_{\l}^m = r {\bf \nabla}Y_{\l}^m,  \quad
{\bf T}_{\l}^m = r \rot {\bf R}_{\l}^m.
\ee
The harmonic decomposition of Eqs. (\ref{eqm_pert}),
(\ref{b_bound}) and (\ref{b_bound1}) is given in appendix.

Equation (\ref{eqm_pert}) reduces to a generalized eigenvalue problem
\st
\be
{\rm [A]}\x=\lambda {\rm [B]} \x,
\ee
where [A] and [B] are differential operators with respect to the $r$ variable only.
The eigenvector associated with $\lambda$ can be written as
\st
\be
\x_{\lambda m}(r)=\left|\begin{array}{l}
\vdots\\
u^\ell_m(r)\\
c^\ell_m(r)\\
w^{\ell+1}_m(r)\\
a^{\ell+1}_m(r)\\
\vdots\\
\end{array}
\right.
\ee
where $\ell$ is running from $m$ to $\infty$.

\subsection{Classification and symmetries}

Because of the axisymmetry of background fields, the different $m$'s of
the spherical harmonic decomposition are not coupled. An additional
simplification comes from the symmetry with respect to equator which
permits the separation of symmetric and antisymmetric modes. A more
detailed discussion of these points can be found in \cite{RRR04} in the
case of pure Alfv\'en modes. In the present case, the Coriolis
acceleration removes the symmetry $m/-m$, which exists for pure Alfv\'en
modes. Moreover, in the axisymmetric case, poloidal and toroidal
components of the fields are coupled.

\begin{figure}
\centerline{
\includegraphics[width=1.0\linewidth,clip=true]{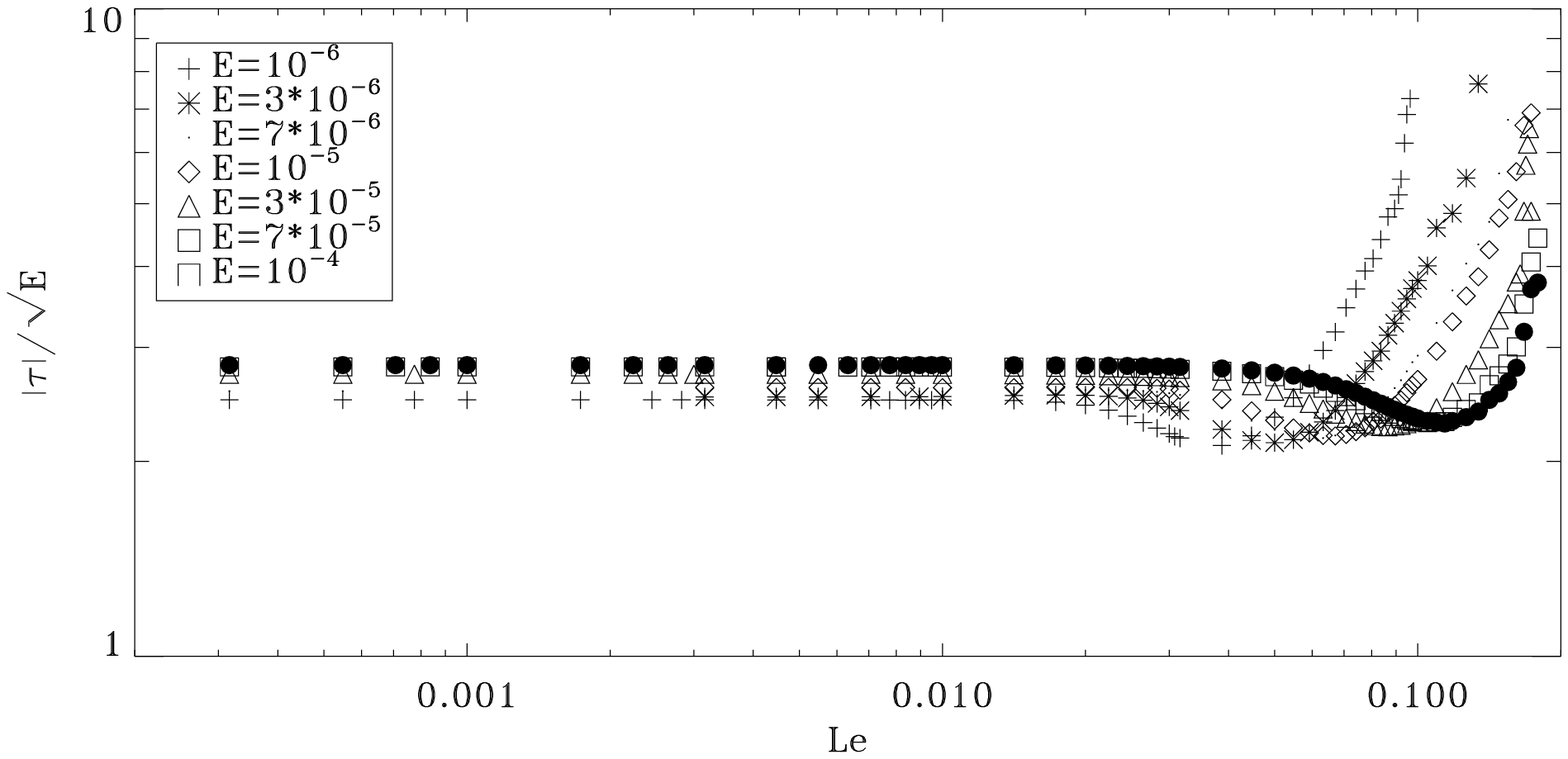}}
\centerline{\includegraphics[width=1.0\linewidth,clip=true]{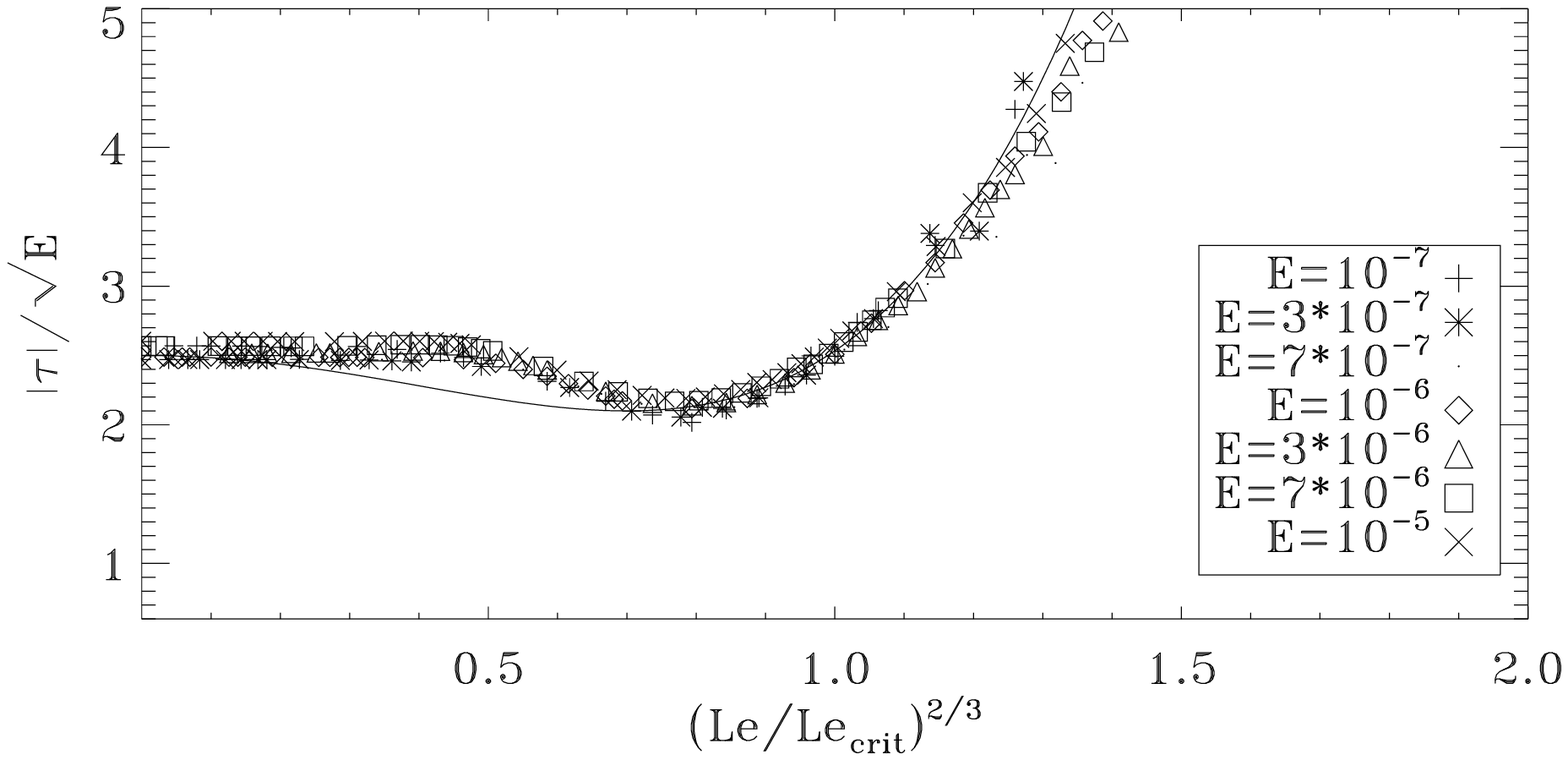}
}
\caption[]{Top : Absolute value of the damping rate of the $m=2$
$r$-mode as a function of the Lehnert number for various Ekman numbers
in $[10^{-7},10^{-4}]$. Bottom: The curves $\tau_E(Le)$ are rescaled by
the critical Lehnert number. The solid line shows the fit
$|\tau|=ax^4+bx^2+2.45$ with $a=1.67, b=-1.63$. In both panels $\eta=0.35$.}

\label{tau_lehn}
\end{figure}

\begin{figure}
\centerline{
\includegraphics[width=1.0\linewidth,clip=true]{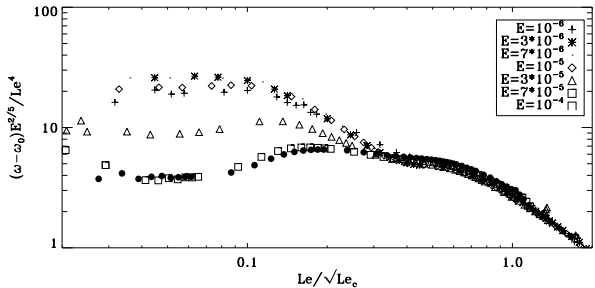}}
\caption[]{Dependence of the frequency of the $m=2$
$r$-mode as a function of the Lehnert number for various Ekman numbers
in $[10^{-6},10^{-4}]$.}
\label{omega_lehn}
\end{figure}

\begin{figure}
\centerline{
\includegraphics[width=1.0\linewidth,clip=true]{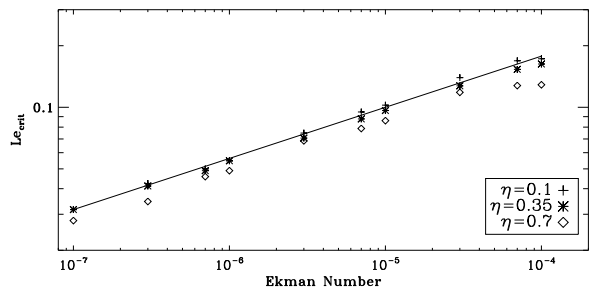}
}
\caption[]{The critical Lehnert number as a function of the Ekman
number for various aspect ratios $\eta$. The straight line shows the $E^{1/4}$-law.}
\label{tau_crit_E}
\end{figure}

\subsection{Numerical aspects}

The numerical resolution of the equations is described in \cite{RRR04}.
We just recall here that the equations governing the radial functions
$\ulm, \vlm, \wlm, \alm, \blm, \clm$ are discretized using the
Gauss-Lobatto grid and the resulting eigenvalue problem is solved either
with a QZ method or with the Arnoldi-Chebyshev algorithm according to
whether we solve for the complete spectrum or a few eigenvalues.

\section{$r$-modes-Alfv\'en waves coupling}

$r$-modes are a subclass of inertial modes which are purely
toroidal. In the case of a non-magnetic and inviscid incompressible
fluid, exact analytical solutions exist \cite[]{RV97}
and the associated velocity field can be written as
\st \be
\u = \alpha \Omega R(r/R)^\l  \Tll,
\ee
where $\alpha$ is an arbitrary constant.  The mode's frequency in the
frame corotating with the fluid is given by
\be
\st
\omega = -2\Omega/(m+1).
\ee
Here, we shall focus on the $m=2$ $r$-mode, which is the most
unstable when coupled to gravitational radiation, and track its
eigenfrequency and damping rate as the magnetic field is increased.

\subsection{The critical Lehnert number}

Since we assumed an infinite magnetic Prandtl number, we noted that the
boundary conditions on the inner core boundary are of no-slip type. This
means that for $\LE\tv0$ the damping rate follows the law derived in
\cite{R01}, namely

\[ \tau=-\frac{35}{2^{7/2}}\frac{1+\eta^6}{1-\eta^7}
{\cal I}_2\sqrt{E}, \quad {\rm with}\quad {\cal I}_2\simeq0.804\]
If $\eta=0.35$, $\tau=-2.494\sqrt{E}$.

In Fig.~\ref{tau_lehn}, we plot the damping rates for various Ekman
and Lehnert numbers. The curves show that when the Lehnert number is
increased, that is when the magnetic field is increased, the (absolute
value of the) damping rate first decreases before rising rapidly.
We define a critical Lehnert number, $Le_c$, such that $|\tau(Le\geq
Le_c)|\geq |\tau(Le=0)|$. When $\tau$ is plotted as a function of the
rescaled Lehnert number, namely $Le/Le_c$, all the curves superpose.

For later use, we fit this curve with a simple polynomial in
$x=(Le/Le_c)^{2/3}$, namely $|\tau|=ax^4+bx^2+2.45$ with $a=1.67, b=-1.63$. The
precise shape of the fit is not crucial as long as the values of the minimum
and the growth beyond it are respected.

In Fig.~\ref{omega_lehn} we show the variations of the frequency of
the $r$-mode with both the Ekman number and the Lehnert number. The
curves have been rescaled so as to show a minimized dependence to the
parameters. This plot suggests some scaling laws for $\omega-\omega_0$,
notably that $\omega\simeq \omega_0+\omega_1\LE^4$, $\omega_1>0$ for weak
magnetic fields. The asymptotic analysis of the behaviour of $\omega$
is quite cumbersome and beyond the scope of this paper.

We show in Fig.~\ref{tau_crit_E} that the critical Lehnert
number varies quite closely as $E^{1/4}$. Actually, a good fit is
$Le_c=1.78E^{1/4}$. Magnetic fields will therefore modify the dynamics
of the oscillations when $Le\geq Le_c$.

\begin{figure}
\centerline{
\includegraphics[width=1.0\linewidth,clip=true]{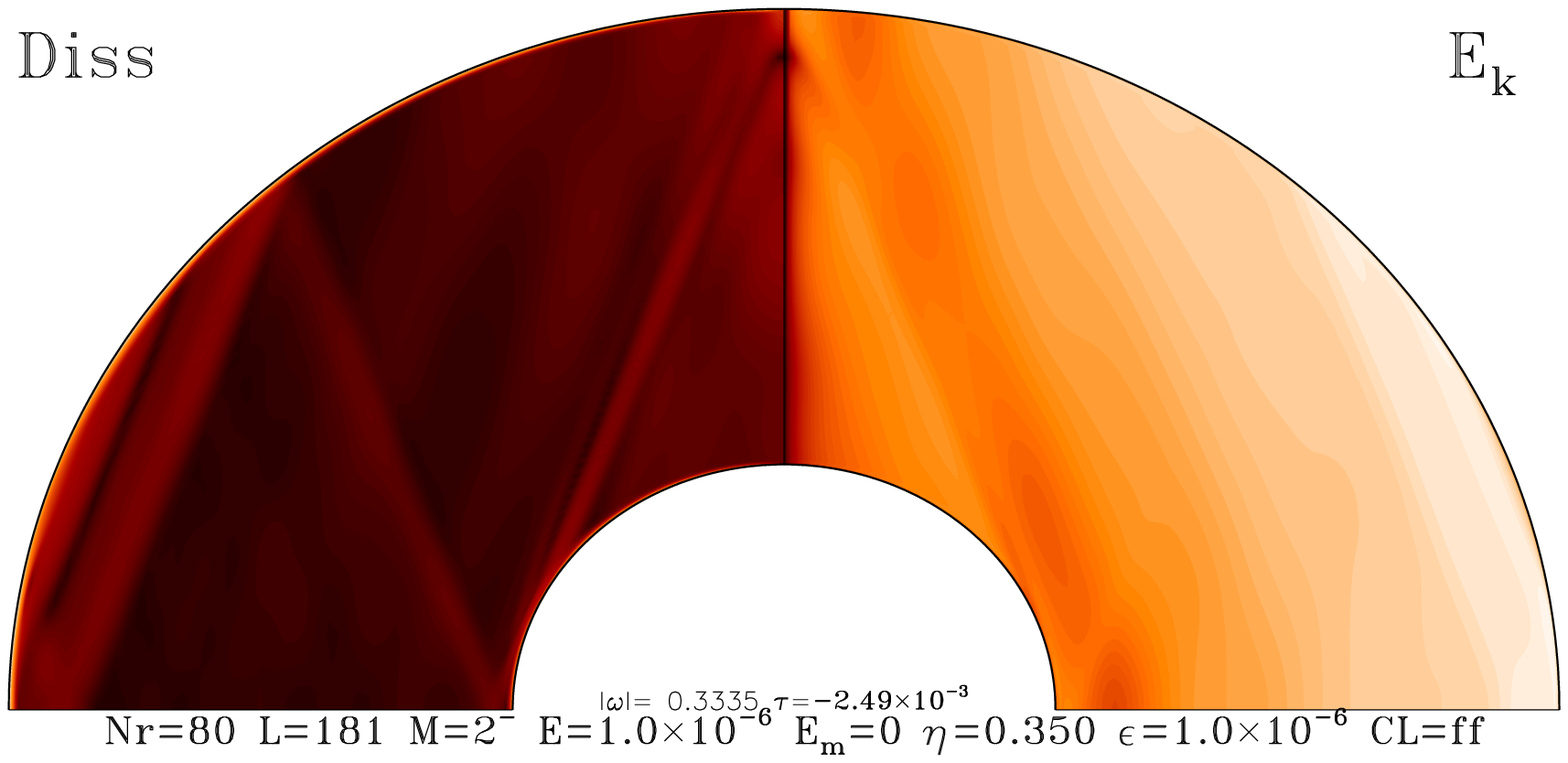}}
\centerline{\includegraphics[width=1.0\linewidth,clip=true]{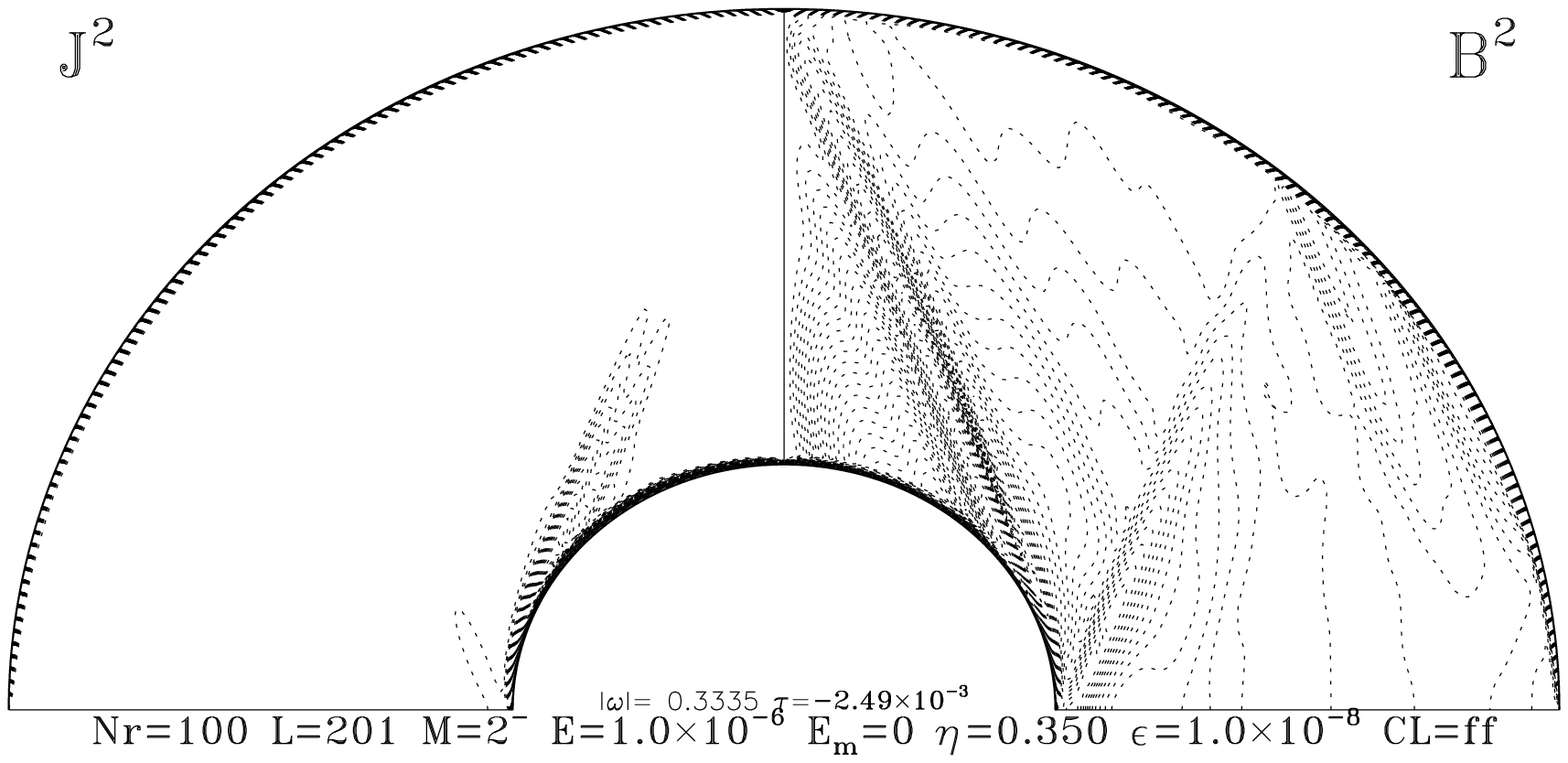}
}
\caption[]{Top : In a meridional plane of the shell, the kinetic energy (right)
and viscous dissipation (left) of the m=2 $r$-mode when a weak
magnetic field is present (small Lehnert number). Bottom: a view of the
associated magnetic perturbation. On right the magnetic energy associated with the
oscillation. Left: the associated current. Here E=10$^{-6}$, Le=10$^{-3}$
and $\eta=0.35$.
}
\label{dvdm}
\end{figure}

The $E^{1/4}$ scaling law of the critical Lehnert number beyond which magnetic
field is influential is rather surprising in view of the discussion of
sect.~\ref{ns_num}. We may understand this scaling if we consider the nature of
$r$-modes when perturbed by a dipolar magnetic field. The magnetic and velocity
perturbations, as long as their gradients, are displayed in Fig.~\ref{dvdm}.
It is clear that the magnetic perturbation is concentrated along the shear
layer emitted at the critical latitude, namely $\arcsin(1/3)=19.5°$. Such shear
layers are a feature of viscous inertial modes which is due to a singularity of
the boundary layer at this latitude \cite[e.g.][]{RV10}.

By analyzing the scale of this layer, as shown in appendix, we find that
when $Le\gg\sqrt{E}$, its thickness scales like $Le/E^{1/4}$, showing
that indeed when $Le$ reaches $E^{1/4}$, the interaction between the
magnetic field and the $r$-mode becomes strong, presumably emphasizing a
resonance between the oscillating shear layer and an Alfv\'en wave. From
the diminishing of the damping rate, we conclude that the dissipative
layers slightly thicken, which is understandable since the magnetic
field tends to oppose to shear.

\begin{figure}
\centerline{
\includegraphics[width=1.0\linewidth,clip=true]{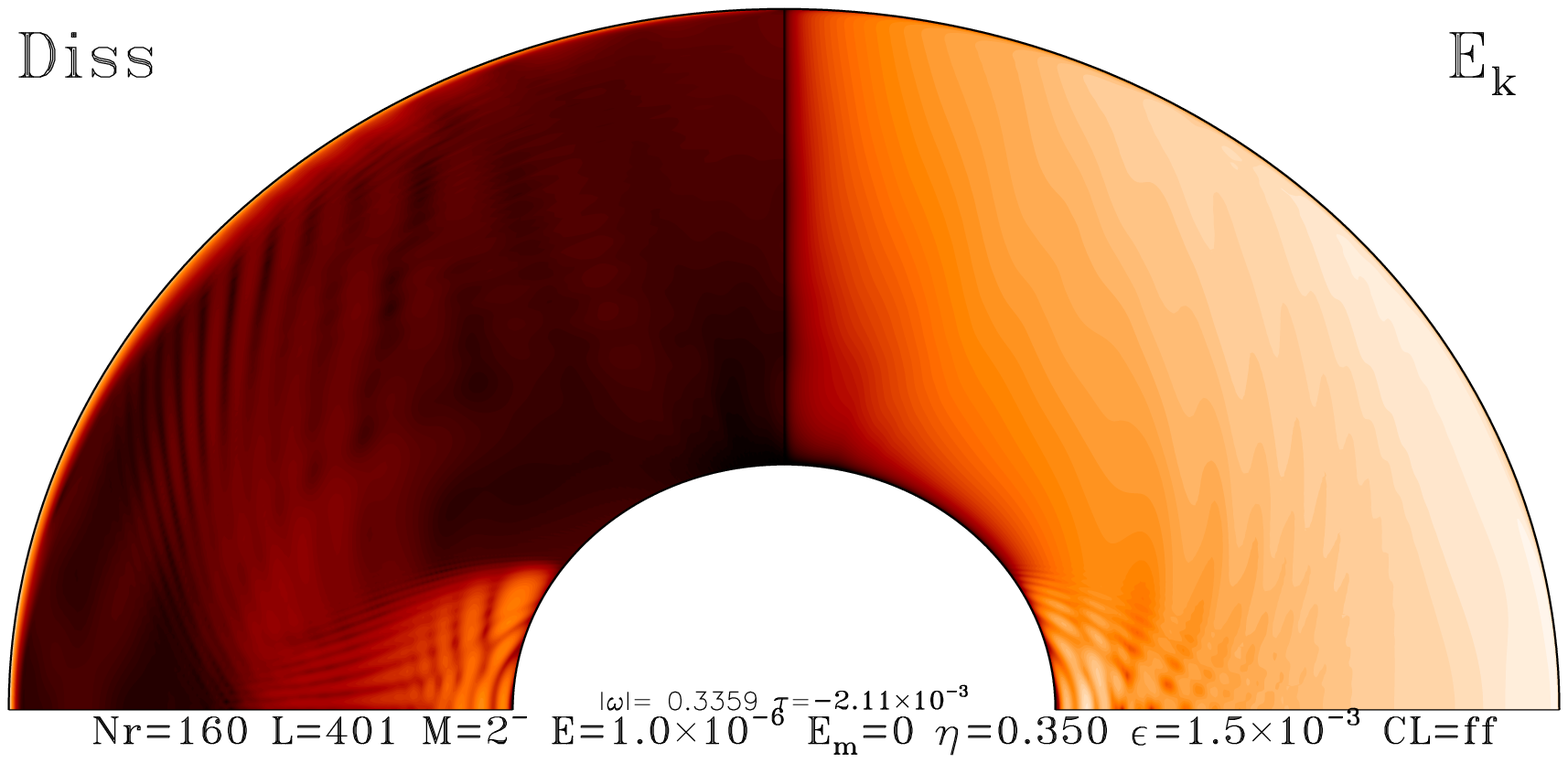}}
\centerline{\includegraphics[width=1.0\linewidth,clip=true]{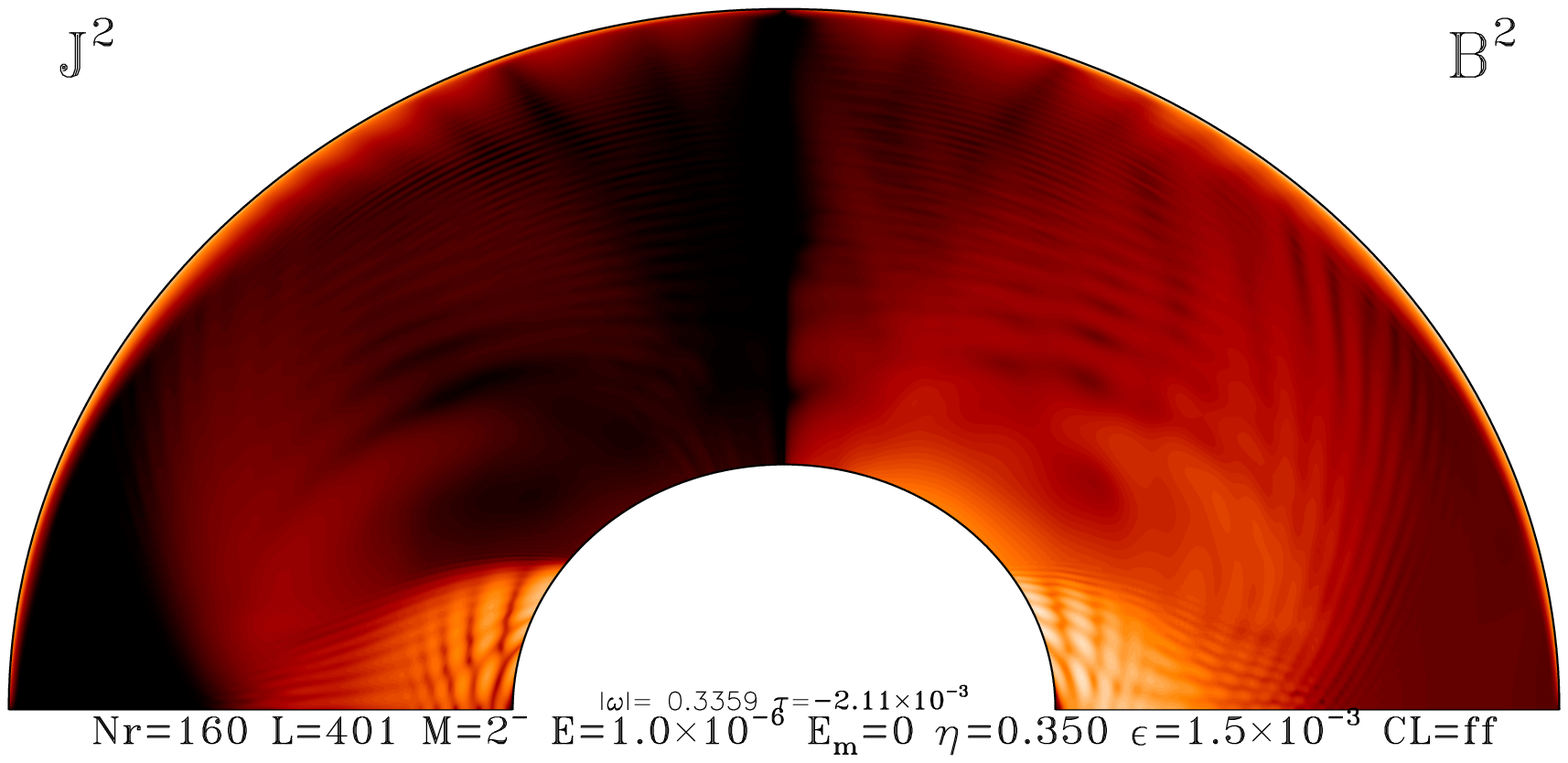}
}
\caption[]{As in figure~\ref{dvdm} but with a critical Lehnert number. Note how the
dipolar magnetic field now features the various part of the oscillation.}
\label{dd_highLe}
\end{figure}

When the Lehnert number is much larger than its critical value the $r$-mode is
completely destroyed, leaving the place to small scale (very dissipative)
Alfv\'en waves (see Fig.~\ref{dd_highLe}).

\subsection{Influence of the magnetic field on the instability}

Let us now turn to the physical consequences of the inequality

\[ Le \geq 1.78E^{1/4}.\]
It may be transformed
into a constraint on the magnetic field, namely

\[ B \geq B_{\rm crit} = 1.78\sqrt{\mu_0\overline{\rho}}2\Omega R E^{1/4}.\]
We now introduce the scaled angular velocity $\Omega_* = \Omega/\sqrt{\pi
G\overline{\rho}}$ and the temperature dependence of the kinematic
viscosity as in \cite{BU00}, namely $\nu = 1.8f/T_8^2$~m$^2$s$^{-1}$,
where $f$ is a parameter of order unity and $T_n=T/10^n$. We thus find

\[ B_{\rm crit} = 7.1\, 10^{14}\frac{{\Omega_*}^{3/4}}{T_9^{1/2}}\;{\rm G}. \]
This critical magnetic field is very high, showing that only hot and
slowly rotating neutron stars might be affected by the magnetic field. To
visualize this effect, it is useful to consider the window of instability
in the plane $(T,\Omega_*)$ as in \cite{R01} or in \cite{BU00}.

In order to calculate the boundaries of the window, we need to approximate
the curves in Fig.~\ref{tau_lehn} with some analytical function. We thus use
the polynomial derived precedingly.

In the plane $(T,\Omega_*)$, the window boundary gives the critical
angular velocity beyond which the $r$-mode instability exists. We therefore
need to solve

\[ \gamma_{gw}+ \gamma_{Bulk} + \gamma_{mi} = 0\]
where we take the growth rate of the mode due to gravitational radiation
$\gamma_{gw}=0.306\;{\rm s}^{-1}\;\Omega_*^6$ from \cite[][]{LOM98}
and the damping by bulk viscosity $\gamma_{Bulk}= -2.2\, 10^{-12}\;{\rm
s}^{-1}\; T_9^6\Omega_\star^2$ from  \cite{LMO99}. These expressions
of $\gamma_{gw}$ and $\gamma_{Bulk}$ have been derived using an
n=1-polytropic model for the neutron star, however we expect that the
density distribution influences the damping rate of the mode with a
factor of order unity,
thus not changing the orders of magnitude of the magnetic fields. We
can thus estimate in a simple manner the critical rotation rate for
various values of the temperature and magnetic field.

\begin{figure}
\centerline{
\includegraphics[width=1.0\linewidth,clip=true]{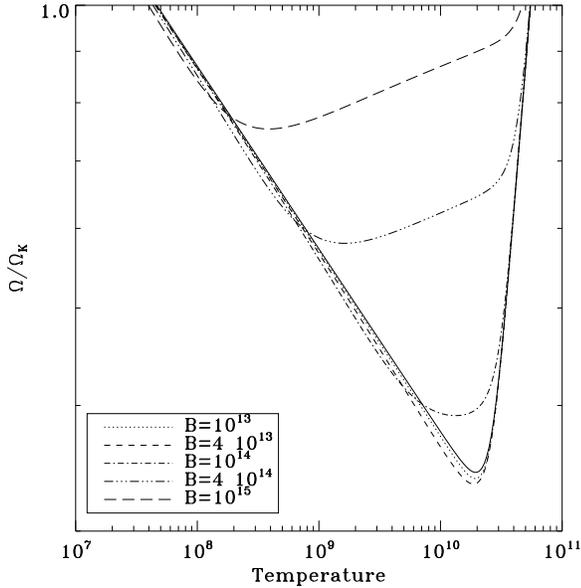}}
\caption[]{The critical angular velocity, normalized by
$\Omega_K=\frac{2}{3}\sqrt{\pi G\overline{\rho}}$ for various values of
the magnetic field. The solid line shows the case with no magnetic field.}
\label{window}
\end{figure}

As expected from the value of the critical magnetic field, we see that
there is a significant reduction of the instability window only when the
field exceeds 10$^{14}$~G, which is rather high. However, we note that
for lower values (e.g. 10$^{13}$ to 4~10$^{13}$~G), the magnetic field
slightly widens the instability window. This is the consequence of the
slight reduction of the damping rate around the critical Lehnert number
that appears in Fig.~\ref{tau_lehn}.

\section{Conclusions}

In this paper we investigated the dynamics of the $m=2$ $r$-mode in a
spherical shell when the fluid is bathed by a dipolar magnetic field. The
main theoretical result of this work is that magnetic fields have a
negligible influence on the $r$-mode unless they exceed a critical
value which is such that the Lehnert number is of the order of the
1/4-power of the Ekman number.  A physical interpretation of this
scaling law may be obtained by observing that at this critical value,
the wavelength of Alfv\'en waves is similar to R$E^{1/4}$ which is
precisely one of the typical thicknesses of shear layers of inertial
modes \cite[e.g.][]{RGV01}.

We also have shown that when the magnetic field is slightly below the
critical value, the damping rate of the mode is slightly reduced because
of the thickening of the various layers due to the frozen field limit.

Back to neutron stars and the famous instability of inertial $r$-modes
when coupled to gravitational radiation, we conclude that, as far as
dipolar fields are concerned, only fields over $10^{14}$ G may seriously
affect the instability. Thus, with a quite different approach,
which includes the specific shear layers of $r$-modes, we can confirm the
result of \cite{Lee05}, obtained with singular perturbations, that for
typical values of $10^{12}$ G, magnetic fields are unimportant.

This result does not therefore invalidate the scenario of \cite{Rez01b},
which suggests that the nonlinear development of the $r$-mode instability
may generate a strong toroidal field from a pre-existing poloidal field.
This mechanism was recently investigated in some details by \cite{CD10}
in the context of accreting neutron stars like low-mass X ray binaries.
Neglecting viscosity, these authors show that magnetic fields above
10$^{12}$~G can reduce the $r$-mode instability. From these results and
ours, we estimate that accounting of viscosity will require fields ten
times stronger, namely above 10$^{13}$~G. However, the details of the
interaction between $r$-modes and a toroidal field in a spherical fluid
layer are poorly known: only a few works \cite[e.g.][and references
therein]{schmitt10} have investigated this question in the context of
the magnetohydrodynamics of the liquid core of the Earth, where the
magnetic Prandtl number is very low. The case relevant to neutron stars
still deserves detailed investigations.

\bigskip
The numerical calculations have been carried out on
the NEC SX8 of the `Institut du D\'eveloppement et des Ressources en
Informatique Scientifique' (IDRIS) and on the CalMip machine of the
`Centre Interuniversitaire de Calcul de Toulouse' (CICT) which are both
gratefully acknowledged.

\appendix
\section{spherical harmonic expansion of the MHD equations}

In this section , we expand each term of Eq. (\ref{eqm_pert}) in the
spherical harmonics in base of ($R_{\l}^m, S_{\l}^m, T_{\l}^m $).
The different parts of the equation have been published elsewhere: The
Lorentz force and the induction equation may be found in \cite{RR03}
while the projection of Coriolis acceleration or its curl are in
\cite{R87,RV97}. For completeness we give here the result of casting
the projection  of these forces into a single equation. The equation of
vorticity yields:

\bea
&&\lambda w^l=\iml w^l + A^l_{l-1}x^{l-1}\dxx \biggl(
\frac{u_m^{l-1}}{x^{l-2}}\biggr) \no \\
&&-A^l_{l+1}x^{-l-2}\dxx\biggl( x^{l+3}u_m^{l+1}\biggr) \no \\
&&+\LE^2\la
\frac{B_r}{\llp r}\lc \Bllm \partial_r(r\clmm)\right.\right. \no \\
&&\left.\left. +\Bllp\partial_r(r\clmp)
-\iml r\Delta_\ell(r\alm)\rc + \right. \no \\
&& \left.\frac{B_\theta}{r}\lc (\ell+1)\allm\clmm -
\ell\allp\clmp\rc \ra+E\Delta_\ell\wlm
\eea

\bea
&&\lambda \Delta_\ell(r\ulm) - \iml \Delta_\ell(r\ulm) + \no \\
&& B^\ell_{\ell-1}r^{\ell-1}\dnr{}
\biggl(\frac{\wlmm}{r^{\ell-1}}\biggr) +
B^\ell_{\ell+1}r^{-\ell-2} \dnr{}\biggl( r^{\ell+2}\wlmp\biggr) \no \\
&& = \LE^2\llp\la\frac{B_\theta}{r}\lc
\ell\Allm\Delta_{\ell-1}(r\almm) \right.\right.\no \\
&&\left.  -(\ell+1)\Allp\Delta_{\ell+1}(r\almp) +im\clm/r \rc  \no \\
&&-B'_\theta im\clm/r \no \\
&&+B_r\lc\Allm {\cal D}_{\ell-1}\almm +\Allp {\cal
D}_{\ell+1}\almp\right. \no \\
&& \left.+ \iml\frac{1}{r}\ddnr{r\clm}\rc \no \\
&& +B'_r\lc\frac{}{} \Allm\Delta_{\ell-1}(r\almm) +
\Allp\Delta_{\ell+1}(r\almp)\right.  +\no \\
&& \left.\left.
\iml\frac{1}{r}\dnr{r\clm}\rc\ra+E\Delta_\ell\Delta_\ell(r\ulm)
\eea
While the induction equation gives:

\begin{eqnarray*}
\lambda r\alm &=& im B_r \wlm + B_\theta\lp (l+1)\allm\ulmm -
l\allp\ulmp\rp +\\
&& B_r\lp B_{l-1}^l\vlmm+B^l_{l+1}\vlmp\rp + E_m
\Delta_\ell(r\alm)\\
\lambda \clm &=& \lc
\frac{B_\theta}{r}\beta^l_k+\frac{(rB_r)'}{r}\gamma^\ell_k\rc\wkm +
B_r\gamma^\ell_k\dnr{\wkm} \\
&& - \iml \lp \llp
B_\theta+ (rB_r)'\rp \frac{\vlm}{r} \\
&& - \iml B_r \dnr{\vlm} +
\frac{(rB_\theta)'}{r}\iml \ulm \\
&&+ \iml B_\theta \dnr{\ulm} + E_m\Delta_\ell\clm
\end{eqnarray*}
and the mass and flux conservation read:

\st
\bea
\stq
&&v_m^\ell=\frac{1}{\ell(\ell+1)r}\dnr{}(r^2u_m^\ell),
\quad \quad\quad (\div \v=0),\\
\stq
&&b_m^\ell=\frac{1}{\ell(\ell+1)r}\dnr{}(r^2a_m^\ell),
\quad\quad\quad(\div \b=0),
\eea
In these equations we introduced the following operators

\[\Delta_\ell=\frac{1}{r}\ddnr{}r - \frac{\llp}{r^2} \]

\[ {\cal D}_\ell = \frac{1}{r}\dnr{}r\Delta_\ell = \frac{1}{r}\dddnr{}r
-
\frac{\llp}{r}\dnr{}\frac{1}{r}\]

and coupling coefficients:

\[\alpha ^l_{l-1}=\alpha _l^{l-1}=\sqrt{\frac{l^2-m^2}{(2l-1)(2l+1)}}\]
\[\beta ^l_{l-1}=(l-1)\alpha ^l_{l-1},\qquad
                 \beta ^l_{l+1}=-(l+2)\alpha ^l_{l+1}\]
\[\gamma_l^k = ( (k(k+1)+l(l+1) -2)/2k(k+1) )\alpha_l^k \]
\[A^l_{l-1}=A_l^{l-1}=\alpha ^l_{l-1}/l^2\]
\[ B^l_{l-1}=B_l^{l-1}=(l^2-1)\alpha ^l_{l-1}\]

All the differential equations need to be completed by boundary
conditions; for the velocity field, no-slip conditions read:

\[\ulm = 0 \qquad \dnr{\ulm}=0\qquad \wlm=0
\]
on both boundaries. These conditions are sufficient when $E_m=0$.

\section{Boundary layer analysis for the critical Lehnert number}

Let us consider cartesian coordinates adapted to the geometry of the shear
layer as shown in Fig.~\ref{dvdm} bottom (as materialized by the region with high
$\vb^2$). $z$ is the coordinate normal to the shear layer assumed
to contain the rapid spatial variations.

The dynamics inside this layer verifies

\[ \lambda\vu + \omega \ez\wedge\vu = -\na p +
\frac{Le^2}{\lambda}\lp\rot\dz{\vu}\rp\times\ez + E\Delta\vu\]
where we assumed that the background magnetic field is along $\ez$ and that the
shear layer makes an angle $\arcsin\omega$ with the rotation axis (hence the
factor $\omega$ in front of Coriolis acceleration). After taking the curl
of this equation, considering only the rapid variations along $\ez$, and taking
its
$x$ and $y$ components, we find

\[ \lambda(\lambda+i\omega)\partial_z\overline{u}=(Le^2-\lambda
E)\partial^3_z\overline{u}\]
with $\overline{u}=u_x+iu_y$.

Noting that for $r$-mode $\lambda=-i\omega+\tau_0\sqrt{E}$, we find that if
$\overline{u}\propto \exp(-z/z_0)$ then

\[ z_0 \sim (1+i) \frac{Le}{E^{1/4}}\]
This result shows that there is a change of scale in the flow when the
Lehnert number reaches values \od{E^{1/4}}.

\bibliographystyle{mn2e}

\begin{thebibliography}{}

\bibitem[\protect\citeauthoryear{Andersson}{Andersson}{1998}]{anderss98}
Andersson N.,  1998, ApJ, 502, 708

\bibitem[\protect\citeauthoryear{{Andersson}}{{Andersson}}{2003}]{And03}
{Andersson} N.,  2003, Classical and Quantum Gravity, 20, 105

\bibitem[\protect\citeauthoryear{{Baym}, {Pethick}, {Pines} \&
  {Ruderman}}{{Baym} et~al.}{1969}]{Bay69}
{Baym} G.,  {Pethick} C.,  {Pines} D.,    {Ruderman} M.,  1969, Nature, 224,
  872

\bibitem[\protect\citeauthoryear{{Bildsten} \& {Ushomirsky}}{{Bildsten} \&
  {Ushomirsky}}{2000}]{BU00}
{Bildsten} L.,  {Ushomirsky} G.,  2000, ApJ, 529, L33

\bibitem[\protect\citeauthoryear{{Bonanno}, {Cuofano}, {Drago}, {Pagliara} \&
  {Schaffner-Bielich}}{{Bonanno} et~al.}{2011}]{BCDPS11}
{Bonanno} L.,  {Cuofano} C.,  {Drago} A.,  {Pagliara} G.,
  {Schaffner-Bielich} J.,  2011, ArXiv e-prints

\bibitem[\protect\citeauthoryear{{Cuofano} \& {Drago}}{{Cuofano} \&
  {Drago}}{2010}]{CD10}
{Cuofano} C.,  {Drago} A.,  2010, Phys. Rev. D, 82, 084027

\bibitem[\protect\citeauthoryear{{Flowers} \& {Itoh}}{{Flowers} \&
  {Itoh}}{1976}]{FI76}
{Flowers} E.,  {Itoh} N.,  1976, ApJ, 206, 218

\bibitem[\protect\citeauthoryear{{Flowers} \& {Itoh}}{{Flowers} \&
  {Itoh}}{1979}]{FI79}
{Flowers} E.,  {Itoh} N.,  1979, ApJ, 230, 847

\bibitem[\protect\citeauthoryear{{Friedman} \& {Morsink}}{{Friedman} \&
  {Morsink}}{1998}]{FM98}
{Friedman} J.~L.,  {Morsink} S.~M.,  1998, ApJ, 502, 714

\bibitem[\protect\citeauthoryear{{Glampedakis} \& {Andersson}}{{Glampedakis} \&
  {Andersson}}{2006}]{GA06}
{Glampedakis} K.,  {Andersson} N.,  2006, MNRAS, 371, 1311

\bibitem[\protect\citeauthoryear{{Haskell} \& {Andersson}}{{Haskell} \&
  {Andersson}}{2010}]{HA10}
{Haskell} B.,  {Andersson} N.,  2010, MNRAS, 408, 1897

\bibitem[\protect\citeauthoryear{{Haskell}, {Andersson} \&
  {Passamonti}}{{Haskell} et~al.}{2009}]{HAP09}
{Haskell} B.,  {Andersson} N.,    {Passamonti} A.,  2009, MNRAS, 397, 1464

\bibitem[\protect\citeauthoryear{{Ho} \& {Lai}}{{Ho} \& {Lai}}{2000}]{HL00}
{Ho} W.~C.~G.,  {Lai} D.,  2000, ApJ, 543, 386

\bibitem[\protect\citeauthoryear{{Jault}}{{Jault}}{2008}]{Jault08}
{Jault} D.,  2008, Phys. Earth Plan. Int., 166, 67

\bibitem[\protect\citeauthoryear{{Kinney} \& {Mendell}}{{Kinney} \&
  {Mendell}}{2003}]{KM03}
{Kinney} J.~B.,  {Mendell} G.,  2003, Phys. Rev. D, 67, 024032

\bibitem[\protect\citeauthoryear{{Lee}}{{Lee}}{2005}]{Lee05}
{Lee} U.,  2005, MNRAS, 357, 97

\bibitem[\protect\citeauthoryear{{Lehnert}}{{Lehnert}}{1954}]{Lehnert54}
{Lehnert} B.,  1954, ApJ, 119, 647

\bibitem[\protect\citeauthoryear{Lindblom, Mendell \& Owen}{Lindblom
  et~al.}{1999}]{LMO99}
Lindblom L.,  Mendell G.,    Owen B.,  1999, Phys. Rev. D, 60, 104014

\bibitem[\protect\citeauthoryear{Lindblom, Owen \& Morsink}{Lindblom
  et~al.}{1998}]{LOM98}
Lindblom L.,  Owen B.,    Morsink S.,  1998, Phys. Rev. Lett., 80, 4843

\bibitem[\protect\citeauthoryear{{Mendell}}{{Mendell}}{2001}]{Men01}
{Mendell} G.,  2001, Phys. Rev. D, 64, 4009

\bibitem[\protect\citeauthoryear{{Reese}, {Rincon} \& {Rieutord}}{{Reese}
  et~al.}{2004}]{RRR04}
{Reese} D.,  {Rincon} F.,    {Rieutord} M.,  2004, A\&A, 427, 279

\bibitem[\protect\citeauthoryear{{Rezzolla}, {Lamb}, {Markovi{\' c}} \&
  {Shapiro}}{{Rezzolla} et~al.}{2001a}]{Rez01a}
{Rezzolla} L.,  {Lamb} F.~K.,  {Markovi{\' c}} D.,    {Shapiro} S.~L.,  2001a,
  Phys. Rev. D, 64, 104013

\bibitem[\protect\citeauthoryear{{Rezzolla}, {Lamb}, {Markovi{\' c}} \&
  {Shapiro}}{{Rezzolla} et~al.}{2001b}]{Rez01b}
{Rezzolla} L.,  {Lamb} F.~K.,  {Markovi{\' c}} D.,    {Shapiro} S.~L.,  2001b,
  Phys. Rev. D, 64, 104014

\bibitem[\protect\citeauthoryear{{Rezzolla}, {Lamb} \& {Shapiro}}{{Rezzolla}
  et~al.}{2000}]{Rez00}
{Rezzolla} L.,  {Lamb} F.~K.,    {Shapiro} S.~L.,  2000, ApJ, 531, L139

\bibitem[\protect\citeauthoryear{Rieutord}{Rieutord}{1987}]{R87}
Rieutord M.,  1987, Geophys. Astrophys. Fluid Dyn., 39, 163

\bibitem[\protect\citeauthoryear{Rieutord}{Rieutord}{1991}]{R91}
Rieutord M.,  1991, Geophys. Astrophys. Fluid Dyn., 59, 185

\bibitem[\protect\citeauthoryear{Rieutord}{Rieutord}{2001}]{R01}
Rieutord M.,  2001, ApJ, 550, 443

\bibitem[\protect\citeauthoryear{Rieutord, Georgeot \& Valdettaro}{Rieutord
  et~al.}{2001}]{RGV01}
Rieutord M.,  Georgeot B.,    Valdettaro L.,  2001, J. Fluid Mech., 435, 103

\bibitem[\protect\citeauthoryear{Rieutord \& Valdettaro}{Rieutord \&
  Valdettaro}{1997}]{RV97}
Rieutord M.,  Valdettaro L.,  1997, J. Fluid Mech., 341, 77

\bibitem[\protect\citeauthoryear{Rieutord \& Valdettaro}{Rieutord \&
  Valdettaro}{2010}]{RV10}
Rieutord M.,  Valdettaro L.,  2010, J. Fluid Mech., 643, 363

\bibitem[\protect\citeauthoryear{{Rincon} \& {Rieutord}}{{Rincon} \&
  {Rieutord}}{2003}]{RR03}
{Rincon} F.,  {Rieutord} M.,  2003, A\&A, 398, 663

\bibitem[\protect\citeauthoryear{Schmitt}{Schmitt}{2010}]{schmitt10}
Schmitt D.,  2010, Geophys. Astrophys. Fluid Dyn., 104, 135

\end{thebibliography}

\label{lastpage}

\end{document}